\documentstyle[11pt,paspconf]{article}
\input{psfig}
\def\mincir{\ \raise -2.truept\hbox{\rlap{\hbox{$\sim$}}\raise5.truept	
\hbox{$<$}\ }}								%
\def\magcir{\ \raise -2.truept\hbox{\rlap{\hbox{$\sim$}}\raise5.truept	%
\hbox{$>$}\ }}								%

\begin{document}

\title{The Rise and Fall of the Quasars} 

\author{A. Cavaliere and V. Vittorini}
\affil{Astrofisica, Dip. Fisica 2a Universit\`a,
    Roma I-00133}

\begin{abstract}
The coherent rise and fall of the quasar population 
is discussed in terms of 
gas accretion onto massive black holes, 
governed by the hierarchically growing 
environment. The rise is related 
to plentiful accretion during the assemblage of the host galaxies; the 
fall to intermittent accretion when these interact  
with companions in a group. 
The LFs are computed out to 
$z= 6$, and are  related to the mass distribution of relict BH 
found in local galaxies. 
The histories of the QS
and of the star light are compared.

\end{abstract}


\vspace*{-.7truecm}
\section{Evidence }

The bright quasars exhibit remarkable permanence and remarkable
variations: the spectra 
(including the optical emission lines) are basically similar for 
individual QSs shining at redshifts in the full range $z \approx 0 - 5$, 
throughout most 
of the universe life; meanwhile, 
 the population undergoes substantial changes. These 
look well organized, as shown by surveys in the radio, the optical 
and the X-ray band 
(see Shaver et al. 1996, Osmer in this Volume, and references therein).
In fact, the bright QS population {\it rises} and {\it falls} steeply,  
culminating at $z \simeq 3\pm 0.5$; 
fig. 1 represents this course in terms of optical light density,
and compares it with the run of the 
star light density as given by Madau 1997.

\begin{figure}[tbh]
\psfig{figure=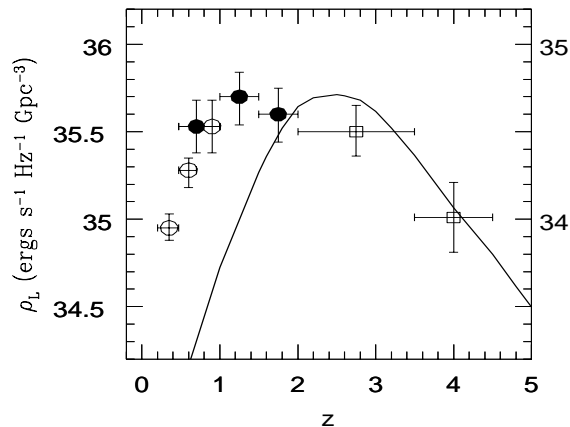,height=5.in,width=6.in}
\vspace*{-6.6cm} 
\caption{The history of the blue luminosity density produced by the QSOs 
with $L_B > 10^{45}$ erg s$^{-1}$ 
is  represented by the solid line 
(scale on the right). The points outline the UV light history of the stars  
after Madau 1997 (left scale). 
For the radio sources, see Dunlop 1997.
\label{fig:fig1}}
\end{figure}

Over the first few Gyrs of the universe life 
the number of bright QSs grows, and the luminosity 
functions mainly rise in normalization, a so-called negative density 
evolution. Instead, after $z \simeq  3$ 
the optical LFs fall to nearly blend in with the 
local Seyfert 1 nuclei. 
Luminosity evolution apparently prevails, but 
some positive density evolution also occurs; 
in addition, the optical LFs flatten toward us  
to comprise more bright objects than previously recognized 
(La Franca \& Cristiani 1997; Goldschmidt \& Miller 1997; 
K\"ohler et al. 1997).

On the other hand, aimed images from HST 
and from the ground in optimal seeing conditions 
(Hutchings \& Neff 1992, 
Disney et al. 1995, Hasinger et al. 1996, Bahcall et al. 1997)
extend to quasars out to $z \simeq  0.3$ 
 the evidence of a complex environment nailed down by Rafanelli et al. 1995 
for the local Seyferts. 
In fact, up to $ 1/2$ of the hosts mapped are found 
to be either engaged in galaxy interactions and in merging events,  
or to have close neighbors, even submerged within the host body.
At a somewhat wider range, the 
QS environments are found to comprise some  $10- 20$ galaxies on average 
(Fisher et al. 1996), 
the membership of a group.

Thus the evidence points toward one basic engine, but with working regimes 
related to the surrounding structure. We shall discuss how such connection 
gives rise to non-monotonic evolution.

\section{Paradigms}

QSs as {\it individual} sources are held to be powered by gas accretion 
onto massive black holes. This had been widely argued to explain 
outputs 
up to $L \sim 10^{48}$ erg s$^{-1}$ with large L/M and high 
apparent compactness, 
considering also the terminal stability of the BHs. 
The paradigm envisages extracting the power $L$ from a mass accretion 
$\dot M$, at distances $r$ down to a few times the Schwarzschild radius 
$r_S$ and with radiative efficiencies up to $\eta = L/c^2 \dot M \sim 
r_S/r \sim 0.1$.

Direct evidence confirming the paradigm is now mounting (see Rees 1997), 
while the need for high $\eta$ is weakening.
This is because 
relict BHs -- expected to 
reside in normal galaxies  
keeping the record of the mass accreted during their 
whole career -- are now found with number $\times$ 
masses exceeding the expectations under top 
efficiency 
(Kormendy \& Richstone 1995, 
Magorrian et al. 1997). 

But the paradigm has little to say 
-- directly -- concerning the 
rise and fall of the QS {\it population}. In fact, 
accretion drawing from an unlimited mass supply 
but self-limited by the BH radiation pressure 
in the Eddington regime 
yields luminosities that scale like $ L_E\simeq 10^{46}\; M_{BH}/10^8\, 
M_{\odot}$ erg s$^{-1}$, and 
exponentiate on the short time scale 
$\eta\, t_E \simeq 4\,  10^7$ yr, or last even less according to Haiman 
\& Loeb 1997. 
Thus the activities of the individual QSs constitute just short {\it flashes}
compared with the evolutionary times 
of a few Gyrs or with the population lifespan of $\sim 12$ Gyr. 
A complementary, 
{\it coordinating} agency is called for, to organize all those flashes 
into a coherent rise and fall over an Hubble time. 

Here enters the other paradigm, the hierarchical growth of structures 
which builds up the BH environment and the gas reservoir, and 
regulates the accretion. 
The gas may be held 
at bay on scales $\magcir 10^2$ pc   in an axisymmetric 
gravitational potential where the angular momentum $j$ is conserved. 
But such symmetry is  broken during the host {\it galaxy} build up, 
when strongly asymmetric events of merging occur; 
these  allow plentiful mass inflow. Subsequently, 
the hosts stabilize  but are enclosed in {\it groups},  
where they interact with companion galaxies. The potential is 
again distorted giving rise to episodes of mass inflow,  
recurring but gradually petering out as groups are reshuffled into clusters. 

We maintain, following up CPV97, that in both dynamical regimes 
$j$ at times is not conserved, 
providing a condition {\it necessary}  
for growing new BHs, or for refueling the old ones. 
The transitional mass from a large {\it galaxy} to a small {\it group} 
is around $5\; 10^{12} \;M_{\odot}$; in the hierarchical cosmogonies  
this corresponds 
to $z \simeq 3 \pm 0.5$, depending on  
cosmogonical and cosmological details. 
In any case, such values interestingly fall in the range where the bright QSs 
peak. 

\section{Bimodal accretion} 

Hierarchical cosmogony (see Peebles 1993) envisages  
larger and larger structures condensing out of 
gravitationally unstable density perturbations dominated by dark matter.  
 In the critical universe 
the typical dark halos condensing and virializing at the epoch $t$, 
so attaining density contrasts  $\rho/\rho_u(z)  \sim 2\,10^2$, 
scale up in mass after 
$M_c \propto t^{4/(n+3)}$; for cold DM 
perturbations $n$ slowly increases with $M$ in the range $
\simeq -2.5 \div -1.5$. The halo growth may be 
visualized as a sequence of merging events of unequal, sometimes comparable, 
blocks.  
Thus  if a typical rich cluster forms now, a small group 
formed at $z \simeq 2.5$, and most galactic bulges formed before. 
But the actual condensed masses 
are widely dispersed around $M_c$; correspondigly, their number 
rises prior to $t_c \propto M^{(n+3)/4}$, but declines   
only slowly thereafter. 
 Even in the adverse open cosmologies with $\Omega_o \ll  1$ 
a similar trend is retained  until 
the perturbations freeze  out at $1+z \simeq 1/\Omega_o$.

A subgalactic building block of DM with  
$M \sim 10^{10}~ M_{\odot}$  
allows a BH of nearly $10^6~M_{\odot}$ to form, involving 
a baryon fraction around $\epsilon \sim  10^{-4}$.
The galaxy assemblage goes on 
 by repeated, chaotic merging; 
the baryons lose angular momentum to the DM at a rate 
set by the number density of  substructures surviving 
in the gas and in the DM, and so approximately proportional to 
$\rho^2_u(z)$. Inflow is plentiful and accretion only self-limited, to yield 
luminosities 
$L\sim L_E \propto M_{BH}  \propto \epsilon \, M$ where  $\epsilon$ 
may still vary with $M$ and $z$ as discussed below. 

In groups, many 
simulations (see Governato, Tozzi \& Cavaliere 1996) have shown 
galaxy interactions to be frequent and 
strong. This is due to the high 
density of galaxies $n_g \propto \rho_u(z)$ 
and to the low velocity dispersion $V \propto 
M_c ^{(1-n)/12}$, still 
so close to the galaxian  $v_g$ as to allow dynamical resonance 
(such conditions no longer hold in clusters). 
Nearly grazing encounters occur frequently  
on the time scale $\tau_r 
\sim 1/ \pi r^2_g \; n_g\;V$, and produce 
outright galaxy aggregations (Cavaliere \& Menci 1997), cannibalism,
 or interactions strong enough to 
perturb the potential and cause $j \neq$ const again. 
Simulations of single, aimed interactions (see Barnes \& Hernquist 1991) 
show in detail how a sizeable fraction of 
the gas in both partners loses $j$ 
 and is driven toward the main galactic nucleus, down to a 
distance limited to now only by the computational dynamic range.

So the BH {\it bimodal} fueling 
during host formation and interactions 
is unified and coordinated by the hierarchical evolution of structures. 
This provides the  necessary external condition   
for mass inflow, while ultimate  acceptance by the BH is set 
by the radiation pressure. All that does not end the story, however. 

Initially, with copious inflow the stability  of any 
intermediate structure 
against energy deposition conceivably limits the BH to $M_{BH} \propto 
(1+z)^{2.5}\, M^{5/3}$ (HNR97), 
corresponding to $\epsilon (M,z) \propto (1+z)^{2.5}\, M^{2/3}$. 
At last, when external conditions allow only inflows 
$\dot M\mincir  10^{-2}L_E/c^2 $, one expects 
energy advection from the accretion disk down the BH horizon 
before the electrons share and radiate the ion energy. 
If so (but see Bisnovatyi-Kogan \&  Lovelace  1997), 
the residual emission should peak in 
radio and in hard X-rays (Di Matteo \& Fabian 1997)
and be low in the optical; 
{\it weak} AGNs at last would break the spectral permanence by 
their non-equilibrium condition.   
This is supported by various lines of evidence: 
optical activity at very low levels 
is detected in a sizeable fraction 
of normal galaxies (Ho, Filippenko \& Sargent 1997); 
X-ray galaxies with narrow optical lines  
are being detected, weak but so numerous as to conceivably saturate 
the XRB at $\sim 10$ keV (see Hasinger 1997); 
these advective accretion flows with low $\eta$ 
may constitute a stealthy addition to the relict BH masses. 

\section{Population kinetics}

Bimodal fueling of BHs is conveniently described 
in terms of two components 
of  the LFs at any $L, z$:
$$N(L,z) = N_1(L,z) + N_2(L,z)~. \eqno(1) $$
The component $N_1$ represents new BHs 
growing and flaring up during the host buildup, 
 and dominates 
for $M < 5\, 10^{12}\, M_{\odot}$ or  for $z \magcir 3$ on average. 
Instead, $N_2$ represents BHs  reactivated by interactions, and dominates in 
structures with $M > 5\, 10^{12}\, M_{\odot}$ for  $z\mincir 3$, 
and especially in groups. 

The computation of $N_1$ 
may be visualized in terms of the kinetic equation 
proposed by Cavaliere, Colafrancesco \& 
Scaramella 1991, which  
(setting the free $f(M)$ from the integration so as to conserve 
the total condensed mass) 
generates the mass fuction 
$N(M,z)$ of Press \& Schechter  1974:
$$\partial_t N=N/\tau_+ -N/\tau _- ~~~~~~~~ with ~~\tau _- = 3t/2,~~~ 
\tau_+ = 
\tau _- (M/M_c)^{-(n+3)/3}~. \eqno (2)$$
The positive driving term 
describes the build up of new host halos on a time scale $t_{g} \sim t$. 
Their destruction on a similar time scale is 
described by the negative term. 
BHs masses and luminosities 
grow in the self-limited regime for $\Delta t \sim 
\eta \, t_E \mincir t$
mainly by such merging events 
rather than by continuos accretion onto a standing BH.  
Then the LFs are obtained in the form 
$N_1(L,z) \, dL \propto  (\Delta t / t)\, \rho_u^2(z)\, N(M,z) \,  dM$. 
 
A more complex story concerns the component $N_2$ (see fig. 2). 
This is driven 
on the time scale $\tau_r$ by the random 
reactivations of the $N_r$ dormant BHs, 
so the driving term for the kinetics is here $N_r/ \tau_r$. 
\begin{figure}[tbh]
\vskip 0.1cm
\psfig{figure=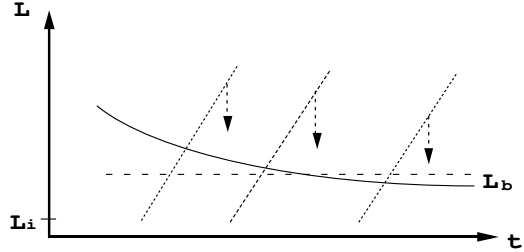,height=1.5in,width=3.8in}
\vspace*{-.2cm} 
\caption{An outline of the individual luminosities from 
intermittent interactions. Solid line: the average $L_b$ 
for accretion of gas in the host; dashed line: 
accretion of gas from satellites. \label{fig:fig2}}
\end{figure}
Now the activity is 
{\it supply-limited} and mostly 
sub-Eddington (Cavaliere et al. 1988, Small \& Blandford 
1992). The BHs restart 
their bright career from a low luminosity 
$L_i$, and brighten up  
on the scale of 
the flyby time $ 2r_g/V \sim r_g/v_g \sim 2\, 10^8$ yr,
which is taken care of by 
the transport term  $ \partial_L(\dot L N_2)$ in the kinetics. 
They attain an average $L_b$, which sets the break in the LFs; 
statistically they quench off with increasing probability $\tau_L^{-1} \propto 
(L/L_b)^{\phi}\,r_g/v_g$ (with $\phi \sim 1$), and 
this sets the slope at the bright end.   
All that is 
expressed by the kinetic equation 
$$\partial_t N_2+\partial_L(\dot{L} N_2)= \delta (L-L_i)\, N_r/\tau _r-
N_2/ \tau_L~.  \eqno(3)$$

To close the argument, the last term integrated over $L$ provides one input 
to the number $N_r$ of dormant BHs, the other coming from those still 
forming and flashing up as described by $N_1$; the component 
$N_2$ arises in groups from 
$N_1$ and requires no independent normalization. We normalize $N_1$ 
to the data at $z \simeq 4$. Adopting the stability limit    
$M_{BH} \propto  (M/10^{13}M_{\odot})^{2/3} M$ (specifically, we use 
the coefficient $10^{-3}\,[(1+z)/5]^{2.5}$)  
has interesting 
virtues (HNR 1997): 
adequately flat  LFs obtain at high-$z$; 
the number of QSs  visible for 
$\Delta t \sim 0.1\, t_E$ is consistent with 
that of the high-$z$, star-forming galaxies (Steidel et al. 1996). 

\section{High and low-$z$ luminosity functions} 

Fig. 3a shows the LFs for $z \magcir 3$ computed 
in the  critical universe using for the perturbations
the tilted CDM spectrum normalized to {\it COBE/DMR} (see Bunn \& White 1997); 
these are 
compared  with optical data adopting the bolometric factor $\kappa =10$.
For such redshifts not enough groups have yet formed for $N_2$ to 
emerge to relevance; 
the evolution of $N_1$ looks like the negative DE type. 

For $z \mincir 3$, instead, $N_2$ emerges and dominates 
the evolution.
But to actually compute it, we have to consider 
where the main gas reservoir resides. 
Fig. 3b represents the low-$z$ 
behavior when the accreted gas is provided mainly by the {\it host} reservoir 
(say, with a constant fraction used up in each interaction, 
also constituting the main gas sink); 
then $-\dot M_{gas}/M_{gas} 
\simeq \tau_r^{-1} \simeq -\dot L_b/L_b$ holds.
But in groups or clusters evolving hierarchically 
in the  critical universe $\tau_r \propto t$ closely obtains, 
since $V 
\propto t$ obtains for $n\simeq -2$, 
while $n_g \propto \rho_u (z)\propto t^{-2}$ applies. 
The result is 
$L_b(t) \propto t^ {-t_o/\tau_{ro}}$; 
with $\tau_{ro}$ around  $6$ Gyr, scaled to groups 
from the classic census of local interacting galaxies by Toomre 1977, 
this reads $L_b \propto (1+z)^{3}$. 

This implies LE dominant for $z < 3$, 
with LFs flattened at the faint end by the  
brightening over the flyby time, 
$N(L) \rightarrow L^{-1 - \tau/\tau_r(t)}$.
But fig. 3b shows that quite some DE also 
occurs, due to the 
decreasing duty cycle of the reactivations governed by $\tau_r(t) \propto t$.
The LFs are steep at the bright end, $N(L) \rightarrow exp\, 
[(-L/L_b)^{\phi}/\phi]$;
but at low $z$ the flat and slow $N_1 (L,z)$  remains exposed  
and flattens the overall slope, not unlike the data 
referenced in Sect. 1. 

\begin{figure}[tbh]
\psfig{figure=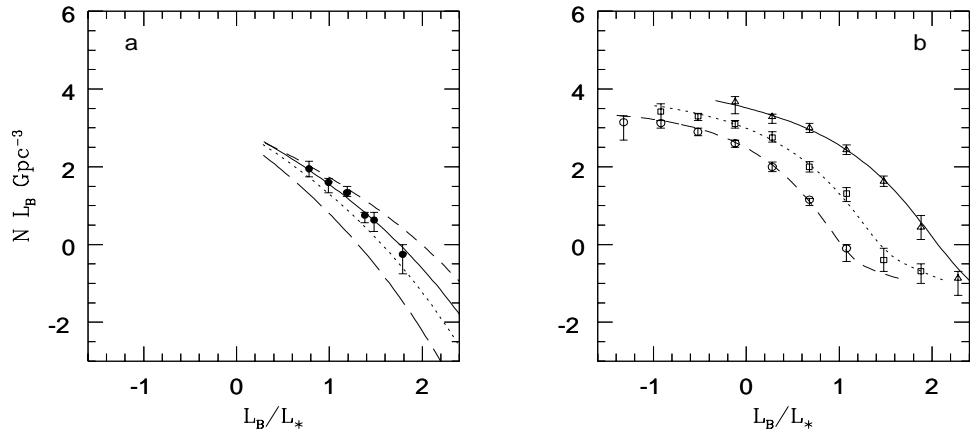,height=5.in,width=5.5in}
\vspace*{-6.6cm} 
\caption{Optical luminosity functions ($\kappa = 10, \, L_* = 10^{45}$ 
erg s$^{-1}$). Panel {\it a} from bottom: 
$z$ = 6, 5, 4.3, 3.5. Panel {\it b} from top: $z$ = 2.5, 1, 0.5. 
Data: Kennefick, Djorgovski \& de Carvalho 1995; Schmidt, Schneider \& 
Gunn 1995; Boyle, Shanks \& Peterson 1988.
\label{fig:fig3}}
\end{figure}

Fig. 4b outlines the mass distribution of the relict BHs   
expected in most 
normal galaxies close to us, computed  from $M_{BH} = \int dt\, L(t)/\eta c^2$ 
with the luminosities evolving as above. 

\begin{figure}[tbh]
\psfig{figure=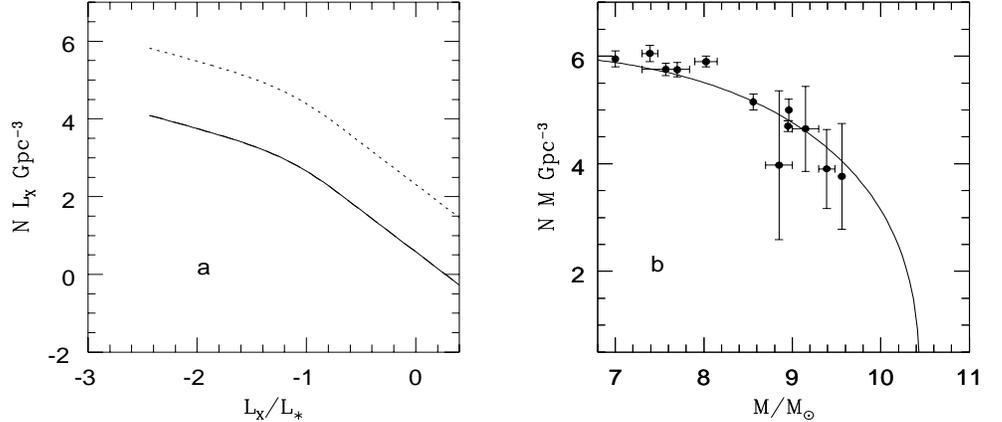,height=5.in,width=5.5in}
\vspace*{-6.6cm} 
\caption{{\it a)} X-ray LFs 
from accretion of gas in satellite 
galaxies, 
with $\eta/\kappa = 10^{-4}$; from top $z=  1, 0.1$.  
{\it b)} Mass distribution 
of the relict BHs, compared with data: masses from 
Kormendy \& Richstone 1995, and numbers from the bulge distribution 
after Franceschini et al. 1998.
\label{fig:fig4}}
\vspace*{-.6cm}
\end{figure}

When the gas in the host is depleted, the remaining reservoir 
is in {\it satellite} galaxies cannibalized out of an initial 
retinue gradually used up. With this prevailing,  
$L_b \sim$ const obtains with no LE (see fig. 2), 
while the DE is enhanced 
 as shown in fig. 4a. The gas available per event $M_{gas} \propto M_{sat}$  
follows the dwarf galaxies distribution, to 
yield steeper LFs at the faint end. We expect such to be the case 
when $\dot M \mincir 10^{-2} L_E/ c^2$ holds, with ADAF prevailing;
we relate this regime to the X-ray LFs of the NLXGs as given by Hasinger 1997. 

\section{Conclusions and discussion}

We conclude that 
the coherent, non-monotonic QS evolution, 
with its scales of a few Gyrs and overall duration 
for $\sim 12$ Gyr, calls for a {\it coordinating} role 
of the surrounding 
structures to govern the accretion onto massive BHs. 

Remarkably, this is provided by the monotonic hierarchical cosmogony. 
Young host galaxies are assembled from subgalactic units, and 
then the mature hosts are packed into 
larger and larger groups where they still evolve for a while 
by interactions. 
Both the assemblage and the interactions  
perturb the host gravitational potential and cause, besides starbursts, 
first rapid growth of BHs, then accretion episodes  inevitably 
petering out in rate and strength. 

Thus the QS-galaxy {\it connection} is actually twofold.
At low $z$ much evidence relates AGN and QS hosts to interacting or to 
group environments. At high $z$ one expects 
QSs associated with subgalactic star-forming blocks just looming out; 
one such instance may be the region singled out by Fontana et al. 1997. 

In this view the QS evolution should be marked by number increase as 
$M_c(t)$ marches through the galactic range  up to $5~10^{12}\, 
M_{\odot}$; here the basic scale is 
$t_{g}$. The turning point occurs as 
$M_c(z)$ outgrows  $5~10^{12}\, M_{\odot}$ at $z\simeq 3\pm 0.5$; 
then the number, and even more the luminosities, decrease 
on the stretching scale $\tau_r\propto t$. 
The LFs so computed are 
found to agree in some detail with the 
observations out to $z \simeq 5$, specifically 
for the hierarchy provided by the 
tilted CDM perturbation spectrum in the critical universe. 

The predictions for $z > 5$ look rather  dim; not many bright 
QSs are expected on the basis of the model successful at lower $z$
(see fig. 3a). 
Actually these predictions sensitively depend on 
the two obvious parameters:
the threshold for DM condensations in the Press \& 
Schechter mass function (in fig. 3a we use the canonical  
$\delta_c= 1.69$); 
the baryonic fraction $M_{BH}/M$ packed in BHs 
(this decreases with $M$ like $M^{2/3}$, but is partially balanced 
by the factor $(1+z)^{2.5}$). 
The predictions for low $z$ and weak $L$, instead, depend on whether 
the main gas reservoir is in the host or in satellite galaxies, 
as shown by figs. 3b and 4a. 

The {\it quasar light} history (of gravitational
origin) computed from the above LFs is 
shown in fig. 1, and compared with the {\it star light} history (of 
thermonuclear origin) given by Madau 1997; the latter at $z\mincir 1$ is  
mainly contributed by 
the faint blue galaxies, which 
 Cavaliere \& Menci 1997 interpret as starbursts in dwarfs interacting in 
large-scale structures. Then the similarity of the two histories 
at such $z$ reflects the basically similar run of the time scale 
$\tau_r \propto n^{-1}_g$ for {\it interactions} in dense environments,   
in spite of two differences: the environments 
are constituted by condensing LSS or by virialized 
groups, respectively; the FBG starbursts last longer than the nuclear 
activities. 

Looking briefly at other hierarchical 
cosmogonies, the hot + cold DM perturbations --  
even when given all advantages like only 20\% hot matter and 
suitable amplitude to fit the data at $z \simeq  4$ -- 
produce  at low $z$ far too many bright QSs, reflecting the 
later collapses of galaxies in this version of the hierarchy. 
In very open universes with standard CDM perturbations, 
we find for $z<2$ too few bright QSs, 
due to  group formation frozen long before and 
to gas reservoirs already exhausted. 
Low-density but flat universes 
fare much better, nearly as the critical, tilted CDM case. 





\end{document}